\documentclass[aps,prb,reprint,superscriptaddress]{revtex4-2}

\usepackage{amssymb}
\usepackage{graphicx}

\usepackage[pdfusetitle,
 bookmarks=true,bookmarksnumbered=false,bookmarksopen=false,
 breaklinks=false,pdfborder={0 0 0},pdfborderstyle={},backref=false,colorlinks=true]
{hyperref}

\begin{document}
\title{A first-principles study of theoretical superconductivity on RbH by doping without applied pressure}

\author{S. Villa-Cort{\'e}s}
\affiliation{Instituto de F{\'i}sica, Benem{\'e}rita Universidad Aut{\'o}noma de Puebla, Apartado Postal J-48, 72570, Puebla, Puebla, M{\'e}xico }

\author{M. A. Olea-Amezcua}
\email{monica.olea@correo.buap.mx}
\affiliation{Benem{\'e}rita Universidad Aut{\'o}noma de Puebla, Escuela de Artes Plásticas y Audiovisuales, Vía Atlixcáyotl No. 2499, 72810, Puebla, Pue., M{\'e}xico }

\author{O. De la Pe{\~n}a-Seaman}
\affiliation{Instituto de F{\'i}sica, Benem{\'e}rita Universidad Aut{\'o}noma de Puebla, Apartado Postal J-48, 72570, Puebla, Puebla, M{\'e}xico }

\date{\today}

\begin{abstract}
The structural, electronic, lattice dynamics, electron-phonon coupling, and superconducting properties of the alkali-metal hydride RbH, metalized through electron-doping by the construction of the solid-solution Rb$_{1-x}$Sr$_x$H,
are systematically analyzed as a function of Sr-content within the framework of density functional perturbation and Migdal-Eliashberg theories, taking into account the effect of zero-point energy contribution by the quasi-harmonic approximation. 
For the entire studied range of Sr-content, steady increments of the electron-phonon coupling constant and the superconducting critical temperature are found with progressive alkaline-earth metal content through electron-doping, reaching the values of $\lambda=1.92$ and $T_c=51.3(66.1)$~K with $\mu^*$=0.1(0). 
The steady rise of such quantities as a function of Sr-content is consequence of the metallization of the hydride as an increase of density of states at the Fermi level is observed, as well as the softening of the phonon spectrum, mainly coming from H-optical modes. Our results indicate that electron-doping on metal-hydrides is an encouraging alternative to look for superconductivity without applied pressure. 
\end{abstract}


\maketitle

\section{Introduction}

In the last years, experimental findings on high critical temperature superconducting (SC) metal hydrides have triggered a significant number of investigations around their outstanding phonon-mediated properties to find optimal superconducting materials. Based on first-principles calculations, together with the Migdal-Eliashberg theory \cite{Zhang2017,eliashberg1960}, a variety of metallic hydrides have been proposed theoretically as candidates for room-temperature superconductors under high applied pressures; as the case of metallic hydrogen \cite{Ash1968, Clay2015} or the widely studied hydride LaH$_{10}$, which presents superconductivity with a experimental critical temperature, $T_{\rm c}$, of around 250-260 K at pressures of about 170-200 GPa \cite{Drozdov2019,Soma2019}.

Furthermore, it is well known that a trustworthy criterion to reach a high $T_{\rm c}$ in hydride materials is a specific combination of strong covalent bonding between hydrogen and other elements of the compound, and the presence of high-frequency modes in the phonon spectrum \cite{Drozdov2015}. On that wise, theoretical calculations have allowed the prediction of the electronic, dynamic and electron-phonon (el-ph) coupling properties of hydrogen-rich compounds at high pressures; leading to promising SC systems with $T_{\rm c}$ near room temperature \cite{villa2022,Zhang2021,Tsuppayakornaek2021,Duan2018,SUKMAS2020156434,SUKMAS2022163524,Pinsook_2020}. One of them concluded that in the observed $T_{\rm c}$ evolution as a function of pressure, the role of the energy-dependence on the density of states at the Fermi level, $N(0)$, is crucial for the proper analysis and description of the superconducting state on high-$T_{\rm c}$ metal hydrides as H$_3$S, by considering its van Hove singularities \cite{villa2022}. H$_3$S exhibits the experimental measure of conventional SC at $T_{\rm c}$=203 K under a hydrostatic pressure of 155 GPa \cite{Drozdov2015}, the understanding of the physical phenomena and mechanism involved in its SC behavior is of utmost importance.

Meanwhile, other theoretical publications report metallic hydride SC properties without applied pressure by the formation of alloys and solid solutions \cite{Olea2019, Villa2021}.
In a previous work, the el-ph coupling and superconductivity of the alkali-metal hydrides LiH, NaH, and KH, metallized through doping with the alkaline-earth metals Be, Mg, and Ca, respectively, were studied by first-principles methods and the Migdal-Eliashberg theory \cite{Olea2019,eliashberg1960}. 
The alkali-metal hydride family, at atmospheric conditions, crystallizes in the NaCl (B1) structure, adopting a stoichiometry MH, with M = Li, Na, K, Rb, or Cs. The band gap of this group is quite large and varies between 4 and 6 eV \cite{van2007}. In accordance with preceding {\it ab initio} calculations, the gap of these pristine hydrides closes between 300 and 1000 GPa \cite{Hooper2012, Lebgue2003}.
However, the resulting band overlap occurs because of induced pressure, and is unlikely to generate a $N(0)$ high enough, indicating poor prospect for them to be good candidates for high-$T_{\rm c}$ superconductors. 
Nevertheless, substituting an alkali-metal with an alkaline-earth metal, the dopant acts as a donor which delivers electrons to the system, obtaining a $n$-doped material. 
Thus leading a steady increase of the el-ph coupling constant $\lambda$ with progressive alkaline-earth metal doping for the already mentioned (Li/Be)H, (Na/Mg)H and (K/Ca)H systems, as a result of two effects: the softening of the phonon spectrum, mainly of the H-optical modes, and the increment of $N(0)$. 
Then, $T_{\rm c}$ can reach values of 2.1 K for Li$_{0.95}$Be$_{0.05}$H, 28 K for Na$_{0.8}$Mg$_{0.2}$H, and even 49 K for K$_{0.55}$Ca$_{0.45}$H \cite{Olea2019}, demonstrating that doping is an alternative route to reach high $T_{\rm c}$ in this hydride class without the need to apply external pressure.

The exploration of new hydride alloys and solid solutions, according to theoretical investigations, could provide an additional and complementary direction to applied pressure to metallize hydride systems, thereby efficiently assisting the design of target SC materials. Hence, in this paper, we present a systematic study of the structural and electronic properties, as well as the lattice dynamics, el-ph coupling, and superconductivity of the alkali-metal hydride RbH doped with alkaline-earth metal Sr, within the framework of density functional theory (DFT) \cite{Kohn} and the self-consistent virtual crystal approximation (VCA) \cite{Omar2009}. 
The crystal structure was optimized within the quasi-harmonic approximation (QHA)\cite{baroni}, at several concentrations, obtaining for each of them their electronic band structure and density of states.  
Meanwhile, the lattice-dynamical properties were computed using density functional perturbation theory (DFPT) \cite{Baroni2001, Heid1999, Baroni1991}. These quantities allow the evaluation of the microscopic el-ph interaction, like the Eliashberg function $\alpha^{2}F(\omega)$, which is required as an input to the strong-coupling Migdal-Eliashberg theory \cite{eliashberg1960}. 
The current results suggest that the metallization of the RbH by electron-doping could be an attractive bypath to achieve high-temperature superconductivity under accessible conditions.

\section{Methodology}

The structural and electronic ground-state properties of the pristine RbH and Rb$_{1-x}$Sr$_{x}$H solid solutions were obtained by {\it ab initio} calculations under the scheme of DFT \cite{Kohn}, meanwhile the lattice dynamics and el-ph coupling properties were carried by DFPT \cite{Baroni2001,Heid1999,Baroni1991}, implemented in the QUANTUM ESPRESSO suite code \cite{0953-8984-21-39-395502}. Exchange and correlation contributions were taken into account in the generalized gradient approximation (GGA), by the Perdew–Burke–Ernzerhof (PBE) functional \cite{PBE}.
The energy cut-off was 60 Ry for the calculation of all properties and 240 Ry for the charge density, while a Monkhorst–Pack $24\times24\times24$ $k$-point mesh was employed for the Brillouin zone (BZ) integration, with a Gaussian smearing of 0.02 Ry \cite{fu}.

A $8\times8\times8$ $q$-point mesh was used to determine the phonon spectra through a Fourier interpolation of the calculated dynamical matrices.
In addition, the zero-point energy (ZPE) effects are taking into account by QHA \cite{Waller1956} in order to include corrections to the ground-state and lattice dynamical properties due to quantum fluctuations at zero temperature. In this way, all the properties analyzed along the present manuscript (structural, electronic, lattice dynamics, and el-ph) incorporate ZPE corrections.

We used VCA \cite{Omar2009} to simulate the solid solutions. This approximation allows to partially substitute one constituent element of the original compound by another one, that is a nearest neighbor in the periodic table, at a given concentration $x$. The ionic potential is represented by the pseudopotential generated for a virtual atom with fractional nuclear charge, which provides a realistic representation of the average electronic charge density in the system as it happens in the real one, while remaining the symmetry of the crystal structure unchanged. 
The implementation of the VCA in QUANTUM ESPRESSO \cite{0953-8984-21-39-395502} averages the local potentials, while wave function integrals over the projectors of the constituent pseudopotentials are evaluated individually. To form the virtual atoms we used weighted averages of the Rb and Sr UltraSoft PseudoPotentials (USPP) as constructed by A. Dal Corso \cite{DAL}, and were generated in scalar-relativistic mode with projectors for the $s$, $p$, and $d$ atomic orbitals. The local potentials take into account the all-electron (AE) potential smoothing through Bessel functions, by a cutoff radius of 1.5 and 1.9, for Sr and Rb, respectively. 

Additionally, in order to analyze the SC properties, the perturbative method provides a route to study the microscopic screened el-ph matrix elements $g^{{\bf q}j}_{{\bf k}+{\bf q}\nu',{\bf k}\nu}$ \cite{Olea2019}, which are employed in the strong-coupling Eliashberg theory \cite{eliashberg1960}. These matrix elements, calculated over a denser $48\times48\times48$ $k$-point mesh, describe the scattering of an electron from a Bloch state with momentum ${\bf k}\nu$ to another Bloch state ${\bf k}+{\bf q}\nu'$, by a phonon mode ${\bf q}j$  with frequency $\omega_{{\bf q}j}$. 
With such information, the phonon linewidth $\gamma_{{\bf q}j}$ (which arise from the el-ph interaction) can be obtained by the following equation \cite{h3,h1,h2}:

\begin{equation}
\gamma_{{\bf q}j} = 2 \pi \omega_{{\bf q}j} \sum_{{\bf k}\nu\nu'} 
\left| g^{{\bf q}j}_{{\bf k}+{\bf q}\nu',{\bf k}\nu} \right|^{2} 
\delta(\varepsilon_{{\bf k}\nu} - E_{F})
\delta(\varepsilon_{{\bf k}+{\bf q}\nu'} - E_{F}),   
\end{equation}

where $\varepsilon_{{\bf k}\nu}$ represents the one-electron band energies with
momentum $\bf k$ and band index $\nu$, and $E_{F}$ is the Fermi energy. 
Then, with $\gamma_{{\bf q}j}$, it is possible the calculation of the isotropic Eliashberg spectral function, $\alpha^{2}F\left(\omega\right)$, which can be described as: 

\begin{equation}
\alpha^{2}F\left(\omega\right)=\frac{1}{2\pi\hbar N\left(0\right)}\sum_{{\bf q}j}\delta\left(\omega-\omega_{{\bf q}j}\right)\frac{\gamma_{{\bf q}j}}{\omega_{{\bf q}j}},
\end{equation}

where $N\left(0\right)$ is the electronic density of states (per atom and spin) at $\epsilon_{F}$.
To obtain $\alpha^{2}F\left(\omega\right)$, it was required to perform a sum over a denser Fourier interpolated $54\times54\times54$ $q$-point mesh. 
Also, the average electron-phonon coupling constant $\lambda$ \cite{h3,h1,h2,Olea2019} was calculated as:

\begin{equation}
\lambda=2 \int_{0}^{\infty} \frac{\alpha^{2}F(\omega)}{\omega} d\omega 
= \frac{1}{\pi \hbar N(0)}\sum_{{\bf q}j}
\frac{\gamma_{{\bf q}j}}{ \omega_{{\bf q}j}^{2}}.
\end{equation}

Finally, the $T_{\rm c}$ was estimated by numerically solving the Migdal-Eliashberg gap equations on the imaginary axis \cite{eliashberg1960,h4,Bergmann1973,VILLACORTES2018371,omar2010} using the respective $\alpha^{2}F(\omega)$ for each Sr-content ($x$) analyzed, treating the Coulomb pseudopotential as a phenomenological parameter.

The electronic, phonon, el-ph coupling, and SC properties were investigated for the optimized structures of both schemes: with and without ZPE Though, in the present manuscript only the ZPE results are mainly discussed; since the obtained properties over the two schemes were qualitatively very similar, as the main differences lie on the lattice parameters, which give softer phonon spectra for the ZPE scheme.

\section{Results and discussion}
\subsection{Structural properties}

\begin{figure}
\centering
\includegraphics[width=8.4cm]{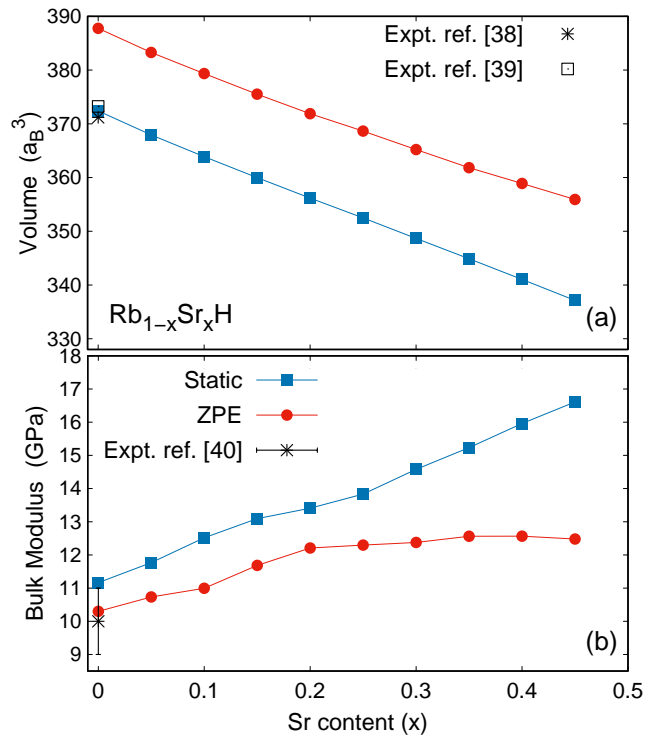}
\caption{(a) Volume (where $a_{B}^{3}=Bohr^3$, atomic units) and (b) bulk modulus ($B_0$) for Rb$_{1-x}$Sr$_{x}$H as a function of Sr-content ($x$). Experimental data obtained from references: \cite{Zin,villars1985,Sangster1994}} 
\label{fig:1}
\end{figure}

At standard pressure values and ambient conditions, the alkali metal Rb combine with hydrogen in a one to one ratio, yielding an ionic solid which crystallizes in the rock-salt structure (B1, space group $Fm\bar{3}m$). Within the VCA, we can assume the same crystal structure for the entire range of Sr-content in the proposed solid solutions. 
We performed structural optimizations for RbH and Rb$_{1-x}$Sr$_{x}$H under two schemes: with (ZPE) and without (static) ZPE corrections at selected concentrations of Sr.
The corresponding ground state volume and bulk modulus (B$_0$) obtained values are resumed in Figure \ref{fig:1}.
The plotted range of Sr-content corresponds to dynamical stable solid solutions, as a result of the phonon dispersion analysis, where it was observed that phonon modes with imaginary frequencies appear for $x \geq$ 0.5.

For the case of pristine RbH, the calculated ground lattice constants are 11.4206 and 11.5756 Bohr for the static and ZPE schemes, respectively, which are in good agreement with the experimental reported value of 11.4309 Bohr \cite{Zin}, with differences of -0.09\% and 1.25\%. In general, taking into account the ZPE contribution drives to a slight unit-cell expansion in the solid solutions of 4.5\% on average, in comparison with the static-energy case, due to the fluctuations around the equilibrium position of the ions caused by their vibrations. The volume decreases in both cases, as the Sr-content increases, meanwhile the volume difference between the two schemes at each concentration gets larger. As discussed in a previous work for other members of the alkali-metal hydrides \cite{Olea2019,Olea2017}, the contraction of the unitary cell can also be attributed to extra-charge redistribution in the interstitial zone as the Sr-content augments in the solid solution. 

Whilst, in relation to the bulk modulus, $B_0$, the experimental value of 10 GPa \cite{Sangster1994} is slightly overestimated by just 3.19\% for the ZPE scheme, whereas the static one shows a difference of 11.17\%.
In both schemes, is notably a continuous increment of $B_0$, with a larger slope for the static case, while for the ZPE one, $B_0$ tends to settle at $x=0.20$. 

\begin{figure}
\centering
\includegraphics[width=8.4cm]{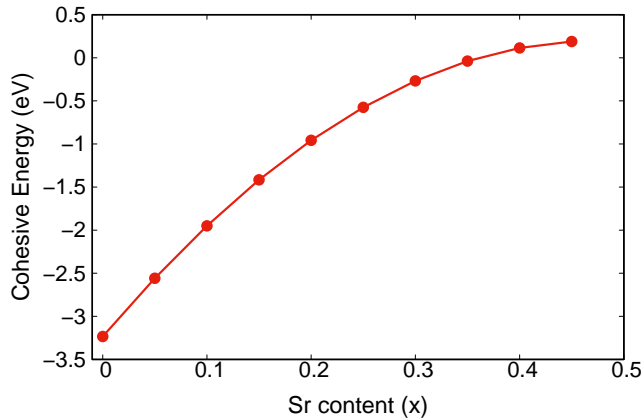}
\caption{Calculated cohesive energy ($E_{coh}$) per formula unit (f.u.) of the Rb$_{1-x}$Sr$_{x}$H solid solution as a function of Sr-content ($x$).} 
\label{fig:2}
\end{figure}

We also have calculated the cohesive energy ($E_{coh}$) for the proposed Sr-contents, $x$, since this quantity allows to characterize the system stability. $E_{coh}$ is obtained by:

\begin{equation}
E_{coh}=E^{tot}_{Rb_{1-x}Sr_{x}H}-(1-x)E^{a}_{Rb}-xE^{a}_{Sr}-E^{a}_{H},
\end{equation}

where $E^{tot}_{Rb_{1-x}Sr_{x}H}$ is the total energy of the solid solution at a selected $x$, while $E^{a}_{Rb}$, $E^{a}_{Sr}$ and $E^{a}_{H}$ are the calculated total energies of the respective isolated atoms. The results are resumed in Figure \ref{fig:2}. It can be seen that the proposed solid solutions are less stable than the pristine RbH, since as the $E_{coh}$ absolute value is larger, more stable is the system. 
Meanwhile, for the cases of $x=0.40$ and $0.45$, we obtained quite small positive $E_{coh}$, indicating possible thermal instability, which could be overcome by careful experimental treatment during material preparation process \cite{Huang2014}. 

\subsection{Electronic properties}

Pristine RbH is an insulator at ambient conditions, and has been reported the experimental measurement of the band gap ($E_g$) of 4.91 eV at zero pressure, albeit at 120 GPa, it shrinks to 2.68 eV \cite{gap}. Our calculations throw a direct $E_g$ value of 2.62 eV at the high symmetry $L$-point, for both schemes, which, even when is underestimated by the current DFT methods, it is in good agreement with previous theoretical reports: 2.20, 2.94 and 2.96 eV \cite{gap1,gap2,gap3}.

\begin{figure}
\centering
\includegraphics[width=8.4cm]{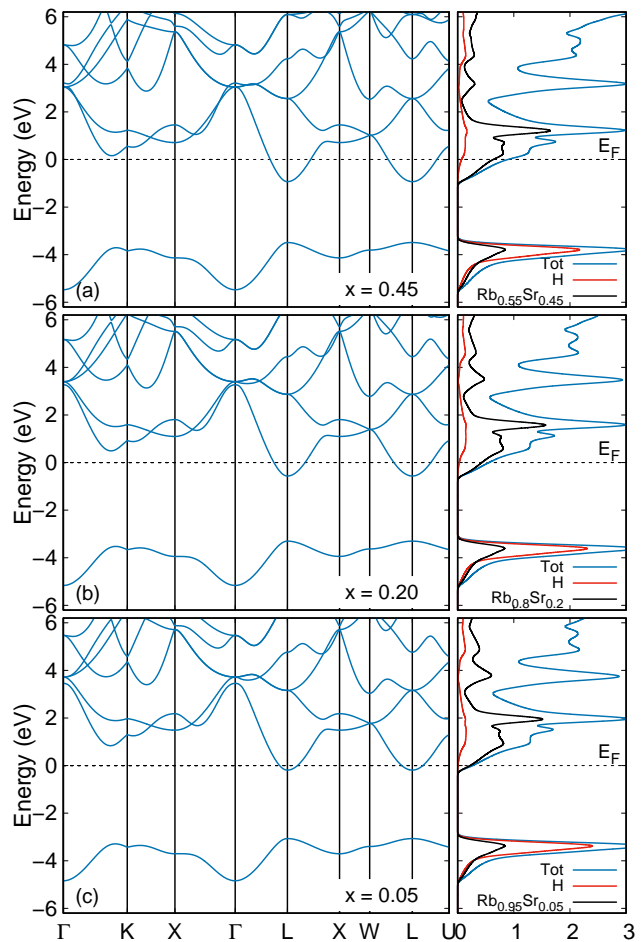}
\caption{Electronic band structure, calculated with the ZPE scheme, for Rb$_{1-x}$Sr$_{x}$ with  $x=0.05$, $0.20$ and $0.45$, including their respective DOS and PDOS.} 
\label{fig:3}
\end{figure}

In order to analyze the evolution of electronic properties as the Sr-content increases, we present in Figure \ref{fig:3} the electronic band structure along high-symmetry paths in the first Brillouin zone and the corresponding density of states (DOS) and partial density of states, calculated within the ZPE scheme, for specific Sr-content cases: $x=0.05$, $0.20$ and $0.45$, to schematize the representative behavior of the results. 
In general, taking into account the ZPE in the calculations does not affect the electronic properties in a significant way, since it only produces a slight reduction of conduction band energies at energy ranges far away from $\epsilon_{F}$, which is expected being that, generally, larger volumes induce smaller bandwidths. 
The valence-band regions are completely filled, mainly formed by H states and a small contribution of Rb states. As is expected, as the Sr-content is raised, the system becomes metallic, since the extra electrons provided by the alkaline-earth element fill states at the conduction band, leading to a metallic behavior without the need to apply external pressure. 
This creates an ellipsoidal Fermi surface centered at the $L$-point, which is slightly anisotropic in $k$-space, with electron-character and formed mainly by Rb/Sr states. 

\subsection{Lattice dynamics}

\begin{figure}
\centering
\includegraphics[width=8.4cm]{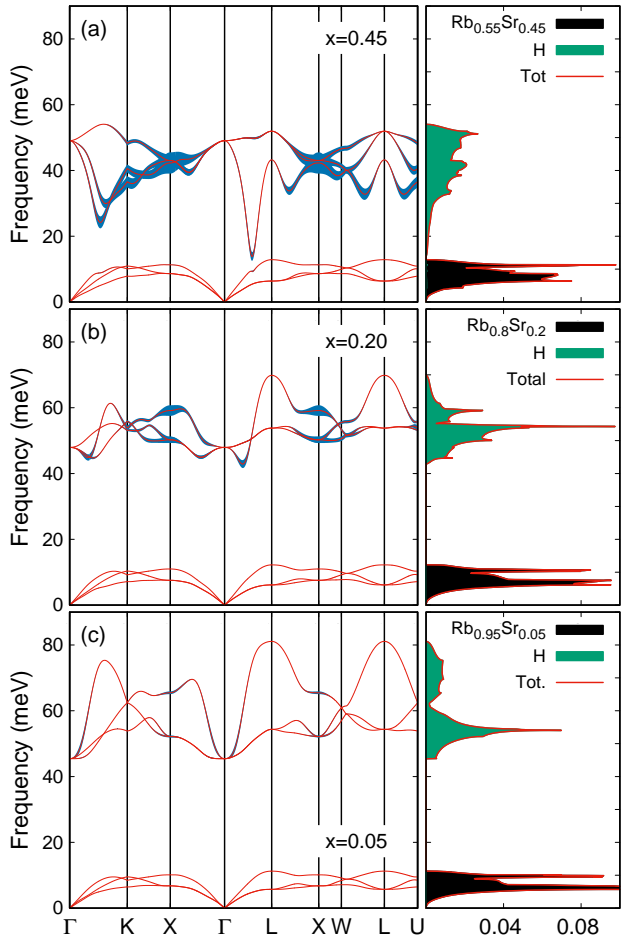}
\caption{Phonon dispersion, including linewidths as verticals lines, and their respective PDOS, calculated with the ZPE scheme for Rb$_{1-x}$Sr$_{x}$H at $x=0.05$, $0.20$, and $0.45$.} 
\label{fig:4}
\end{figure}

With the aim to find the range of Sr-content under which the proposed solid solution Rb$_{1-x}$Sr$_{x}$H remains dynamically stable, the phonon dispersion for $0 \leq x \leq 0.5$ were calculated. As mentioned before, modes with imaginary frequencies were observed when $x=0.5$ (not showed here), demonstrating that 45\% of Sr-content is the maximum to ensure theoretical dynamic stability. In this way, the phonon dispersions, including their respective phonon linewidths ($\gamma_{{\bf q}j}$) and phonon density of states for the pristine RbH and selected Sr-content ($x=0.05$, $0.20$ and $0.45$) are displayed in Figure \ref{fig:4} (under the ZPE scheme). 
In general, the phonon spectra for the ZPE scheme are slightly softer than the static scheme results (not showed here) for the entire range of studied Sr-content.
This softening is linked to the expanded unit cell (see Figure \ref{fig:1}) and is less notable for the acoustical branches in comparison with the optical branches. 

Overall, the optical-frequencies region presents, three well defined branches related to H phonon modes, which is completely separated from the acoustic one, where the modes are mainly related to the metallic atoms.
This is also verified by their respective PDOS. 
Such separation is partly due to the large mass difference between the hydrogen and the metallic atoms. 
It can be seen from Figure \ref{fig:4} that the optical region phonon spectrum softens as $x$ increases, whereas the acoustical part remains rather unchanged with only slight increments in its frequencies. This can be related to the enhanced metallicity in the solid solution, since the electronic density redistribution leads to an increased screening that reduces the hydrogen vibrational frequencies \cite{Olea2017}. 
Another feature, as the Sr-content augments, is the emergence of phonon anomalies in the optical branches at specific regions in the Brillouin zone: $\Gamma-K$, $\Gamma-L$, and in a minor degree at approximately half-way of the $L-X$, $W-L$ and $L-U$ high-simmetry paths.
These are the same anomalies which induce dynamical instabilities as the Sr-content gets closer to $x = 0.5$. Phonon softening and instabilities in metal hydrides induced by alloying have been also observed in previous reports \cite{Zeng2012,Song}.

Additionally, Figure \ref{fig:4} also shows a pronounced increment of the linewidths  ($\gamma_{{\bf q}j}$) as $x$ grows, represented by vertical blue lines along the phonon branches, indicating a possible enhancement of the el-ph coupling on the system \cite{line}.
Larger $\gamma_{{\bf q}j}$ values are located mainly around the $X$-point over the optical H phonon modes.
The coupling of these phonons with large wave vectors as well as the observed phonon anomalies essentially involves electronic states on distinct electron pockets (inter-pocket coupling) \cite{Olea2019}.
These results are in good agreement with other studies of high-$T_{\rm c}$ hydride superconductors, where the optical hydrogen high-frequency phonon modes were found to be responsible for the large el-ph coupling \cite{Drozdov2015,Shamp,Syed}.

\subsection{Electron-phonon coupling and superconducting properties}

\begin{figure}
\centering
\includegraphics[width=8.4cm]{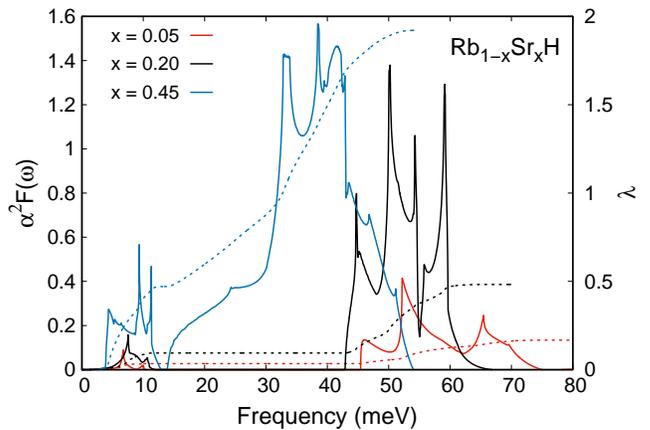}
\caption{Eliashberg spectral function $\alpha^{2}F(\omega)$ and el-ph coupling constant $\lambda(\omega)$ for Rb$_{1-x}$Sr$_{x}$H at $x=0.05$, $0.20$ and $0.45$, calculated under the ZPE scheme.} 
\label{fig:5}
\end{figure}

Finally, we developed the analysis of the el-ph coupling and superconducting properties of the proposed solid solutions within the Eliashberg formalism \cite{eliashberg1960}. 
The Eliashberg spectral functions $\alpha^{2}F(\omega)$ (solid lines) and el-ph coupling parameters $\lambda(\omega)$ (dashed lines), calculated by the partial integration of $\alpha^{2}F(\omega)$, for specific Sr-content values ($x=0.05$, $0.20$ and $0.45$) are drawn in Figure \ref{fig:5}.
For the case of $\alpha^{2}F(\omega)$, we observe that the function shifts to lower frequencies with the rise of alkaline-earth metal content, meantime, its total weight heightens, mostly in the optical high-frequency region. Both changes enhance the el-ph coupling, resulting in a notable growth of $\lambda(\omega)$ as the Sr-content increases.

\begin{figure}
\centering
\includegraphics[width=8.4cm]{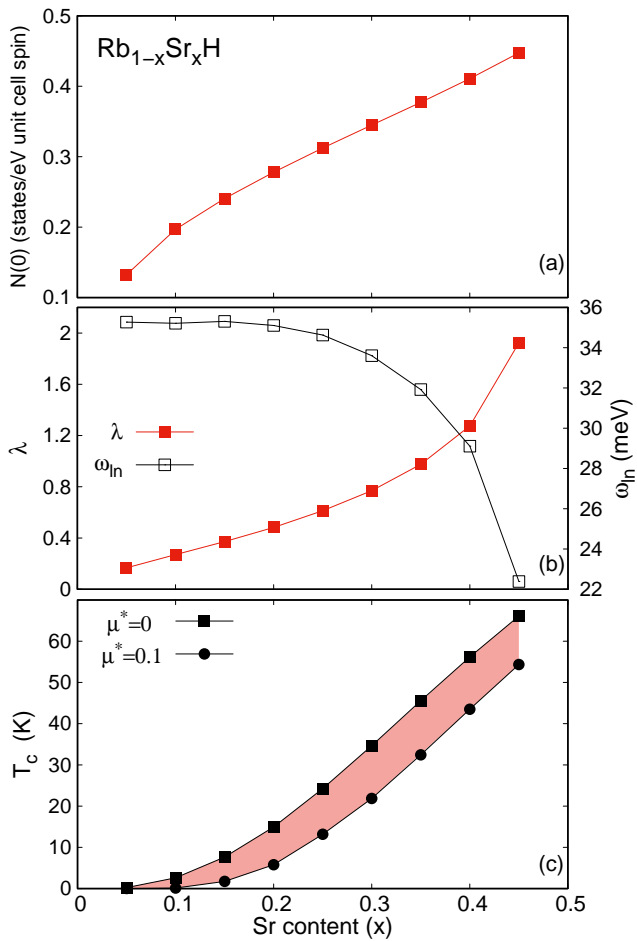}
\caption{Total density of states at the Fermi level {\it N}($0$), average el-ph coupling constant $\lambda$, Allen-Dynes characteristic phonon frequency $\omega_{ln}$, and superconducting critical temperature $T_{\rm c}$ as a function of content $x$ for the Rb$_{1-x}$Sr$_{x}$H alloy.} 
\label{fig:6}
\end{figure}

To achieve a complete understanding of the solid solution superconductivity, it is required the analysis of three important properties that steer the el-ph coupling constant $\lambda$: the density of states at the Fermi level {\it N}($0$), the phonon frequencies $\omega$, and the el-ph coupling matrix elements, considered by the linewidths \cite{h3,h1,h2,Olea2019}. In this way, in Figure \ref{fig:6}, we present the calculated {\it N}($0$), $\lambda$, and the Allen-Dynes characteristic phonon frequency $\omega_{ln}$ \cite{dynes}, in order to examine the role of these factors, on the superconducting critical temperature $T_{\rm c}$ \cite{eliashberg1960,h4,h5,h6,omar2010}, plotted in the same image, for the entire proposed Sr-content range.

Firstly, it can be observed in Figure \ref{fig:6}(a) that {\it N}($0$) steadly grows as $x$ increases, as a result of the electron-doping increment by the incorporation on the alkaline-earth metal in the solid solution. Such {\it N}($0$) rise should, as a first approximation, improve the superconducting properties of the system, a hypothesis that is confirmed by the forthcoming analysis.
Meanwhile, for the case of the average effective frequency $\omega_{ln}$ (see Figure \ref{fig:6}(b)), as is expected from the lattice dynamics discussion, it drastically reduces as $x$ grows, indicating a $\lambda$ enhancement, and possibly also for $T_{\rm c}$, since the low frequency region tends to boost the el-ph coupling, as can be seen from Eq. (3).
Then also in Figure \ref{fig:6}(b), $\lambda$ shows a steady increment for low Sr-content $x$, while the grow-ratio starts to increase as $x$ also increases, as expected, reaching a maximum value of 1.92 at the threshold  $x=0.45$ content.
This particular behavior as mentioned before, has its roots on the performance of both, $N(0)$ and frequencies (described by $\omega_{ln}$) as a function of Sr-content.
At last, for the numerical solution of the isotropic Migdal-Eliashberg gap equations in
order to get $T_{\rm c}$, two different values of the Coulomb pseudopotential $\mu^*$ were employed: 0 and 0.1. The value $\mu^*=0$ provides an upper limit for $T_{\rm c}$, while $\mu^*=0.1$ gives a more realistic estimation for it.
In accordance to the previous analyses, the predicted $T_{\rm c}$ values, in both cases, exhibit a steady increase as a function of $x$, achieving the maximum values of 66.1 K ($\mu^*=0$) and 54.34 K ($\mu^*=0.1$) for $x=0.45$. Despite the fact these values could be considered modest compared with the obtained for H$_3$S at an hydrostatic pressure of 155 GPa (around 203 K), the metallization of this hydride by electron-doping could be an attractive alternative route to achieve high-temperature superconductivity under accessible conditions for other related systems as well.
In addition, comparing with the expected maximum $T_{\rm c}$ for previous studied doped alkali metal hydrides, with 2.1 K, 28 K and 49 K for (Li/Be)H, (Na/Mg)H, and (K/Ca)H, respectively \cite{Olea2019}, it can be confirmed the observed trend of rising $T_{\rm c}$ as the mass of the metal increases.

\section{Conclusions}
A systematic study of the structural, electronic, lattice dynamics, el-ph coupling, and superconducting properties of the proposed solid-solution Rb$_{1-x}$Sr$_{x}$H was accomplished by first principle calculations, taking into account 
the zero-point energy effects on all of them.
From electronic structure analysis, we noticed an increment of the density of states at the Fermi level as the Sr-content increases, demonstrating the metallization of the hydride. From the phonon dispersions of the solid-solution, we found that the continuous incorporation of Sr induces a softening of the phonon spectrum and the emergence of phonon anomalies, mainly coming from the H optical modes, leading to dynamical instabilities for $x \geq$ 0.5. 
The microscopic origin of these instabilities could be related to the formation of ellipsoidal Fermi surfaces centered at the high symmetry $L$-point as the hydride becomes metallic. 
With respect to the el-ph coupling and superconducting properties, the growth of the average el-ph coupling constant with electron-doping, reaching a maximum value of $\lambda=1.92$ at $x=0.45$, is the result of mainly two effects: the steady increment of the density of states at the Fermi level as well as the softening of the phonon frequencies. This leads to a consistent rise of the superconducting critical temperature, with a top value of $54.3(66.1)$~K with
$\mu^*=0.1(0)$ for the Rb$_{0.55}$Sr$_{0.45}$H solid solution. The presented results indicate that metallization of metal hydrides by electron-doping could be an alternative or even complementary route to achieve superconductivity without the need of keeping the system under high applied pressure.

\begin{acknowledgments}
This research was partially suported by the Consejo Nacional de Ciencia y Tecnología (CONACyT, M{\'e}xico) under Grant No. FOP16-2021-01-320399.
The authors thankfully acknowledge computer resources, technical advise, and support provided by Laboratorio Nacional de Superc{\'o}mputo del Sureste de M{\'e}xico (LNS), a member of the CONACyT national laboratories.  
\end{acknowledgments}

\bibliographystyle{unsrt}
\bibliography{sample}

\begin{thebibliography}{10}

\bibitem{Zhang2017}
Lijun Zhang, Yanchao Wang, Jian Lv, and Yanming Ma.
\newblock Materials discovery at high pressures.
\newblock {\em Nature Reviews Materials}, 2(4), February 2017.

\bibitem{eliashberg1960}
GM~Eliashberg.
\newblock Interactions between electrons and lattice vibrations in a
  superconductor.
\newblock {\em Sov. Phys. JETP}, 11(3):696--702, 1960.

\bibitem{Ash1968}
N.~W. Ashcroft.
\newblock Metallic hydrogen: A high-temperature superconductor?
\newblock {\em Phys. Rev. Lett.}, 21:1748--1749, Dec 1968.

\bibitem{Clay2015}
Jeremy McMinis, Raymond~C. Clay, Donghwa Lee, and Miguel~A. Morales.
\newblock Molecular to atomic phase transition in hydrogen under high pressure.
\newblock {\em Phys. Rev. Lett.}, 114:105305, Mar 2015.

\bibitem{Drozdov2019}
A.~P. Drozdov, P.~P. Kong, V.~S. Minkov, S.~P. Besedin, M.~A. Kuzovnikov,
  S.~Mozaffari, L.~Balicas, F.~F. Balakirev, D.~E. Graf, V.~B. Prakapenka,
  E.~Greenberg, D.~A. Knyazev, M.~Tkacz, and M.~I. Eremets.
\newblock Superconductivity at 250 {K} in lanthanum hydride under high
  pressures.
\newblock {\em Nature}, 569(7757):528--531, May 2019.

\bibitem{Soma2019}
Maddury Somayazulu, Muhtar Ahart, Ajay~K. Mishra, Zachary~M. Geballe, Maria
  Baldini, Yue Meng, Viktor~V. Struzhkin, and Russell~J. Hemley.
\newblock Evidence for superconductivity above 260 {K} in lanthanum
  superhydride at megabar pressures.
\newblock {\em Phys. Rev. Lett.}, 122:027001, Jan 2019.

\bibitem{Drozdov2015}
A.~P. Drozdov, M.~I. Eremets, I.~A. Troyan, V.~Ksenofontov, and S.~I. Shylin.
\newblock Conventional superconductivity at 203 kelvin at high pressures in the
  sulfur hydride system.
\newblock {\em Nature}, 525(7567):73--76, August 2015.

\bibitem{villa2022}
S~Villa-Cort{\'e}s and O~De~la Pe{\~n}a-Seaman.
\newblock Effect of van {H}ove singularity on the isotope effect and critical
  temperature of {H}{$_3$}{S} hydride superconductor as a function of pressure.
\newblock {\em Journal of Physics and Chemistry of Solids}, 161:110451, 2022.

\bibitem{Zhang2021}
Xiaohua Zhang, Yaping Zhao, and Guochun Yang.
\newblock Superconducting ternary hydrides under high pressure.
\newblock {\em {WIRE}s Computational Molecular Science}, 12(3):e1582, 2022.

\bibitem{Tsuppayakornaek2021}
Prutthipong Tsuppayakornaek, Wiwittawin Sukmas, Rajeev Ahuja, Wei Luo, and
  Thiti Bovornratanaraks.
\newblock Stabilization and electronic topological transition of hydrogen-rich
  metal {L}i$_{5}${M}o{H}$_{11}$ under high pressures from first-principles
  predictions.
\newblock {\em Scientific Reports}, 11(1), February 2021.

\bibitem{Duan2018}
Defang Duan, Hongyu Yu, Hui Xie, and Tian Cui.
\newblock Ab initio approach and its impact on superconductivity.
\newblock {\em Journal of Superconductivity and Novel Magnetism}, 32(1):53--60,
  October 2018.

\bibitem{SUKMAS2020156434}
Wiwittawin Sukmas, Prutthipong Tsuppayakorn-aek, Udomsilp Pinsook, and Thiti
  Bovornratanaraks.
\newblock Near-room-temperature superconductivity of {Mg}/{Ca} substituted
  metal hexahydride under pressure.
\newblock {\em Journal of Alloys and Compounds}, 849:156434, 2020.

\bibitem{SUKMAS2022163524}
Wiwittawin Sukmas, Prutthipong Tsuppayakorn-aek, Udomsilp Pinsook, Rajeev
  Ahuja, and Thiti Bovornratanaraks.
\newblock Roles of optical phonons and logarithmic profile of electron-phonon
  coupling integration in superconducting {S}c$_{0.5}${Y}c$_{0.5}${H}c$_{6}$
  superhydride under pressures.
\newblock {\em Journal of Alloys and Compounds}, 901:163524, 2022.

\bibitem{Pinsook_2020}
U~Pinsook.
\newblock In search for near-room-temperature superconducting critical
  temperature of metal superhydrides under high pressure: A review.
\newblock {\em Journal of Metals, Materials and Minerals}, 30(2), Jun. 2020.

\bibitem{Olea2019}
M.~A. Olea-Amezcua, O.~De la~Pe{\~{n}}a-Seaman, and R.~Heid.
\newblock Superconductivity by doping in alkali-metal hydrides without applied
  pressure: An ab initio study.
\newblock {\em Physical Review B}, 99(21), June 2019.

\bibitem{Villa2021}
S~Villa-Cort{\'{e}}s and O~De la~Pe{\~{n}}a-Seaman.
\newblock Electron- and hole-doping on {ScH}$_{2}$ and {YH}$_{2}$: effects on
  superconductivity without applied pressure.
\newblock {\em Journal of Physics: Condensed Matter}, 33(42):425401, August
  2021.

\bibitem{van2007}
M.~J. van Setten, V.~A. Popa, G.~A. de~Wijs, and G.~Brocks.
\newblock Electronic structure and optical properties of lightweight metal
  hydrides.
\newblock {\em Phys. Rev. B}, 75:035204, Jan 2007.

\bibitem{Hooper2012}
James Hooper, Pio Baettig, and Eva Zurek.
\newblock Pressure induced structural transitions in {KH}, {RbH}, and {CsH}.
\newblock {\em Journal of Applied Physics}, 111(11):112611, June 2012.

\bibitem{Lebgue2003}
S~Leb{\`{e}}gue, M~Alouani, B~Arnaud, and W.~E Pickett.
\newblock Pressure-induced simultaneous metal-insulator and structural-phase
  transitions in {LiH}: A quasiparticle study.
\newblock {\em Europhysics Letters ({EPL})}, 63(4):562--568, August 2003.

\bibitem{Kohn}
W.~Kohn and L.~J. Sham.
\newblock Self-consistent equations including exchange and correlation effects.
\newblock {\em Phys. Rev.}, 140:A1133--A1138, Nov 1965.

\bibitem{Omar2009}
O.~De~la Pe\~na Seaman, R.~de~Coss, R.~Heid, and K.-P. Bohnen.
\newblock Effects of {A}l and {C} doping on the electronic structure and phonon
  renormalization in {MgB}$_{2}$.
\newblock {\em Phys. Rev. B}, 79:134523, Apr 2009.

\bibitem{baroni}
Stefano Baroni, Paolo Giannozzi, and Eyvaz Isaev.
\newblock {Density-Functional Perturbation Theory for Quasi-Harmonic
  Calculations}.
\newblock {\em Reviews in Mineralogy and Geochemistry}, 71(1):39--57, 01 2010.

\bibitem{Baroni2001}
Stefano Baroni, Stefano de~Gironcoli, Andrea Dal~Corso, and Paolo Giannozzi.
\newblock Phonons and related crystal properties from density-functional
  perturbation theory.
\newblock {\em Rev. Mod. Phys.}, 73:515--562, Jul 2001.

\bibitem{Heid1999}
R.~Heid and K.-P. Bohnen.
\newblock Linear response in a density-functional mixed-basis approach.
\newblock {\em Phys. Rev. B}, 60:R3709--R3712, Aug 1999.

\bibitem{Baroni1991}
Paolo Giannozzi, Stefano de~Gironcoli, Pasquale Pavone, and Stefano Baroni.
\newblock Ab initio calculation of phonon dispersions in semiconductors.
\newblock {\em Phys. Rev. B}, 43:7231--7242, Mar 1991.

\bibitem{0953-8984-21-39-395502}
Paolo~Giannozzi et~al.
\newblock Quantum espresso: a modular and open-source software project for
  quantum simulations of materials.
\newblock {\em Journal of Physics: Condensed Matter}, 21(39):395502, 2009.

\bibitem{PBE}
John~P. Perdew, Kieron Burke, and Matthias Ernzerhof.
\newblock Generalized gradient approximation made simple.
\newblock {\em Phys. Rev. Lett.}, 77:3865--3868, Oct 1996.

\bibitem{fu}
C.~L. Fu and K.~M. Ho.
\newblock First-principles calculation of the equilibrium ground-state
  properties of transition metals: Applications to nb and mo.
\newblock {\em Phys. Rev. B}, 28:5480--5486, Nov 1983.

\bibitem{Waller1956}
Max Born, Kun Huang, and M~Lax.
\newblock Dynamical theory of crystal lattices.
\newblock {\em American Journal of Physics}, 23(7):474--474, 1955.

\bibitem{DAL}
Andrea {Dal Corso}.
\newblock Pseudopotentials periodic table: From {H} to {P}u.
\newblock {\em Computational Materials Science}, 95:337--350, 2014.

\bibitem{h3}
J.~Bardeen, L.~N. Cooper, and J.~R. Schrieffer.
\newblock Theory of superconductivity.
\newblock {\em Phys. Rev.}, 108:1175--1204, Dec 1957.

\bibitem{h1}
Philip~B. Allen.
\newblock Neutron spectroscopy of superconductors.
\newblock {\em Phys. Rev. B}, 6:2577--2579, Oct 1972.

\bibitem{h2}
Philip~B. Allen and Richard Silberglitt.
\newblock Some effects of phonon dynamics on electron lifetime, mass
  renormalization, and superconducting transition temperature.
\newblock {\em Phys. Rev. B}, 9:4733--4741, Jun 1974.

\bibitem{h4}
J.~P. Carbotte.
\newblock Properties of boson-exchange superconductors.
\newblock {\em Rev. Mod. Phys.}, 62:1027--1157, Oct 1990.

\bibitem{Bergmann1973}
G.~Bergmann and D.~Rainer.
\newblock The sensitivity of the transition temperature to changes in
  $\alpha$$_{2}${F}($\omega$).
\newblock {\em Zeitschrift f{\"u}r Physik}, 263(1):59--68, 1973.

\bibitem{VILLACORTES2018371}
S.~Villa-Cort\'es and R.~Baquero.
\newblock The thermodynamics and the inverse isotope effect of superconducting
  palladium hydride compounds under pressure.
\newblock {\em Journal of Physics and Chemistry of Solids}, 123:371 -- 377,
  2018.

\bibitem{omar2010}
O.~De~la Pe\~na Seaman, R.~de~Coss, R.~Heid, and K.-P. Bohnen.
\newblock Electron-phonon coupling and two-band superconductivity of al- and
  c-doped ${\text{mgb}}_{2}$.
\newblock {\em Phys. Rev. B}, 82:224508, Dec 2010.

\bibitem{Zin}
E.~Zintl and A.~Harder.
\newblock Ãber alkalihydride.
\newblock {\em Zeitschrift fÃŒr Physikalische Chemie}, 14B(1):265--284, 1931.

\bibitem{villars1985}
P~Villars and LD~Calvert.
\newblock Pearson's handbook of crystallographic data for intermetallic phases.
  metals park, 1985.

\bibitem{Sangster1994}
J.~Sangster and A.~D. Pelton.
\newblock The h-rb (hydrogen-rubidium) system.
\newblock {\em Journal of Phase Equilibria}, 15(1):87--89, February 1994.

\bibitem{Olea2017}
M~A Olea-Amezcua, J~F Rivas-Silva, O~de~la Pe{\~{n}}a-Seaman, R~Heid, and K~P
  Bohnen.
\newblock Effects of electron doping on the stability of the metal hydride
  {NaH}.
\newblock {\em Journal of Physics: Condensed Matter}, 29(14):145401, March
  2017.

\bibitem{Huang2014}
Juan Huang, Wenhui Xie, and Xiaohong Li.
\newblock The stability, magnetism and electronic structure of
  {F}e$_{15}${TMN}$_{2}$ and {F}e$_{14}${TM}$_{2}${N}$_{2}$ ({TM}={C}r, {M}n,
  {C}o, and {N}i).
\newblock {\em Journal of Magnetism and Magnetic Materials}, 364:1--4,
  September 2014.

\bibitem{gap}
Kouros Ghandehari, Huan Luo, Arthur~L Ruoff, Steven~S Trail, and Francis~J
  Disalvo.
\newblock Crystal structure and band gap of rubidium hydride to 120 {GPa}.
\newblock {\em Modern Physics Letters B}, 9(18):1133--1140, 1995.

\bibitem{gap1}
G.~Sudha Priyanga, A.T.~Asvini Meenaatci, R.~Rajeswara Palanichamy, and
  K.~Iyakutti.
\newblock Structural, electronic and elastic properties of alkali hydrides
  ({MH}: {M}={L}i, {N}a, {K}, {R}b, {C}s): Ab initio study.
\newblock {\em Computational Materials Science}, 84:206--216, 2014.

\bibitem{gap2}
Sinem~Erden Gulebaglan, Emel~Kilit Dogan, Mehmet~Nurullah Secuk, Murat Aycibin,
  Bahattin Erdinc, and Harun Akkus.
\newblock First-principles study on electronic, optic, elastic, dynamic and
  thermodynamic properties of {RbH} compound.
\newblock {\em International Journal for Simulation and Multidisciplinary
  Design Optimization}, 6:A6, 2015.

\bibitem{gap3}
VI~Zinenko and AS~Fedorov.
\newblock First principle calculations of alkali hydride electronic structures.
\newblock {\em Sov. Phys. Solid State}, 36:742, 1994.

\bibitem{Zeng2012}
X.~Q. Zeng, L.~F. Cheng, J.~X. Zou, W.~J. Ding, H.~Y. Tian, and C.~Buckley.
\newblock Influence of 3d transition metals on the stability and electronic
  structure of {MgH}$_{2}$.
\newblock {\em Journal of Applied Physics}, 111(9):093720, May 2012.

\bibitem{Song}
Y.~Song, Z.~X. Guo, and R.~Yang.
\newblock Influence of selected alloying elements on the stability of magnesium
  dihydride for hydrogen storage applications: A first-principles
  investigation.
\newblock {\em Phys. Rev. B}, 69:094205, Mar 2004.

\bibitem{line}
Xiao-Long Zhang and Wu-Ming Liu.
\newblock Electron-phonon coupling and its implication for the superconducting
  topological insulators.
\newblock {\em Scientific Reports}, 5(1), March 2015.

\bibitem{Shamp}
Andrew Shamp and Eva Zurek.
\newblock Superconductivity in hydrides doped with main group elements under
  pressure.
\newblock {\em Novel Superconducting Materials}, 3(1):14--22, 2017.

\bibitem{Syed}
Hasnain~M. Syed, C.J. Webb, and E.~MacA. Gray.
\newblock Hydrogen-modified superconductors: A review.
\newblock {\em Progress in Solid State Chemistry}, 44(1):20--34, 2016.

\bibitem{dynes}
P.~B. Allen and R.~C. Dynes.
\newblock Transition temperature of strong-coupled superconductors reanalyzed.
\newblock {\em Phys. Rev. B}, 12:905--922, Aug 1975.

\bibitem{h5}
Vladimir~Z. Kresin and Stuart~A. Wolf.
\newblock Induced superconducting state and two-gap structure: Application to
  cuprate superconductors and conventional multilayers.
\newblock {\em Phys. Rev. B}, 46:6458--6471, Sep 1992.

\bibitem{h6}
S.~Y. Savrasov, D.~Y. Savrasov, and O.~K. Andersen.
\newblock Linear-response calculations of electron-phonon interactions.
\newblock {\em Phys. Rev. Lett.}, 72:372--375, Jan 1994.

\end{thebibliography}

\end{document}